\documentclass[final,superscriptaddress,english,twocolumn,amssymb,aps,prl,
longbibliography]{revtex4-2}
\usepackage{graphicx}
\usepackage{natbib}
\usepackage{amsmath}
\usepackage{amssymb}
\usepackage{wasysym}
\usepackage{oplotsymbl}
\usepackage{appendix}
\usepackage{soul}
\usepackage[section]{placeins}
\definecolor{darkblue}{rgb}{0,0,.65}
\definecolor{darkgreen}{rgb}{0.28,0.41,0.19}
\definecolor{nicegreen}{rgb}{0.28,0.85,0.19}
\usepackage{lipsum}
\usepackage{nicefrac}

\usepackage[%
  pdfauthor={Benedikt Placke},%
  pdfstartview=FitH,%
  breaklinks=true,%
  bookmarks=true,%
  colorlinks=true,%
  anchorcolor=black,%
  citecolor=darkgreen,
  filecolor=black,%
  menucolor=black,%
  urlcolor=darkblue,%
  linkcolor=blue,%
 ]{hyperref}
\usepackage{diagbox}

\def\equationautorefname~#1\null{Eq. (#1)\null}

\newcommand{\appref}[1]{\hyperref[#1]{App.~\ref*{#1}}}
\usepackage[all]{hypcap} %
\usepackage{physics}

\usepackage[normalem]{ulem}

\begin{document}

\title{Ising Fracton Spin Liquid on the Honeycomb Lattice}

\author{Benedikt Placke}\email{placke@pks.mpg.de}
\author{Owen Benton}\email{benton@pks.mpg.de}
\author{Roderich Moessner}\email{moessner@pks.mpg.de}
\affiliation{Max Planck Institute for the Physics of Complex Systems, Noethnitzer Str. 38, 01187 Dresden, Germany}

\date{\today}

\begin{abstract}
We study a classical Ising model on the honeycomb lattice with local two-body interactions and present strong evidence that at low temperature it realizes a higher-rank Coulomb liquid with fracton excitations.
We show that the excitations are (type-I) fractons, appearing at the corners of membranes of spin flips. Because of the three-fold rotational symmetry of the honeycomb lattice, these membranes can be \emph{locally} combined such that no excitations are created, giving rise to a set of ground states described as a liquid of membranes.
We devise a cluster Monte-Carlo algorithm purposefully designed for this problem that moves pairs of defects, and use it to study the finite-temperature behavior of the model. We show evidence for a first order transition from a high-temperature paramagnet to a low-temperature phase whose correlations precisely match those predicted for a higher-rank Coulomb phase.
\end{abstract} 

\maketitle

One of the central concepts in condensed matter physics is the notion of quasiparticles: the idea that low energy excitations of a system are weakly interacting particle-like objects.
In general, these quasipaticles are capable of independent motion, and it is via this motion that energy inserted locally into the system can spread out, thus allowing equilibration.
Fractons are quasiparticles outside this paradigm, being 
completely immobile when isolated \cite{newman1999glassy, chamon2005fractons, bravyi2011, bravyi2013, castelnovo2011glass, yoshida2013, vijay2015, vijay2016, haah2011code, pretko2020review, nandkoshore2019review}.
Fractons are intimately connected with the conservation laws of exotic gauge theories
involving not only charge but higher moments (e.g. dipole moment) of the charge density \cite{xu2006hrgt, xu2010prd, pretko2017, pretko2017b, Pretko2017gravity}.
It is these conservation laws which render isolated fractons immobile.

Recent years have seen a concerted effort
to establish theoretical models in which
higher moment conservation laws and fracton
physics appear \cite{Pretko2018elasticity, sous2020holes, You2020plaquette, Pretko2019circuits, Hering2021, Yan2022, myerson2022rydberg, giergiel2022bosehubbard, han2022breathing}.
Various lattice models have been proposed to give
rise to fractonic behavior, although these often require
complicated multi-body interactions.
From the point of view of identifying routes to experimental realization, it is preferable to find models built-from short-ranged, two body interactions.

One setting in which the construction of 
such models has been successful is classical
spin systems \cite{yan2020, benton2021topo, zhang2022dynamical}.
Classical models have some
advantages: a Hamiltonian can be readily be constructed to enforce a local constraint of choice in the low energy sector, and this constraint can be chosen in such a way as to reproduce the Gauss's law(s) of a given gauge theory.
Once the model is constructed, it is  in principle accessible
to Monte Carlo simulation.

Thus far these models have been constructed
from continuous degrees of freedom, but this
has the drawback that one cannot isolate and study discrete fractons. Moreover, it is difficult to reintroduce quantum fluctuations in a controlled way. If fracton physics could be demonstrated in a classical Ising model, fractons will naturally be present as discrete excitations, and there exists a clear route to studying quantum effects by introducing off-diagonal terms, e.g. via a transverse field.

%==========
%-----------

\begin{figure}
\includegraphics{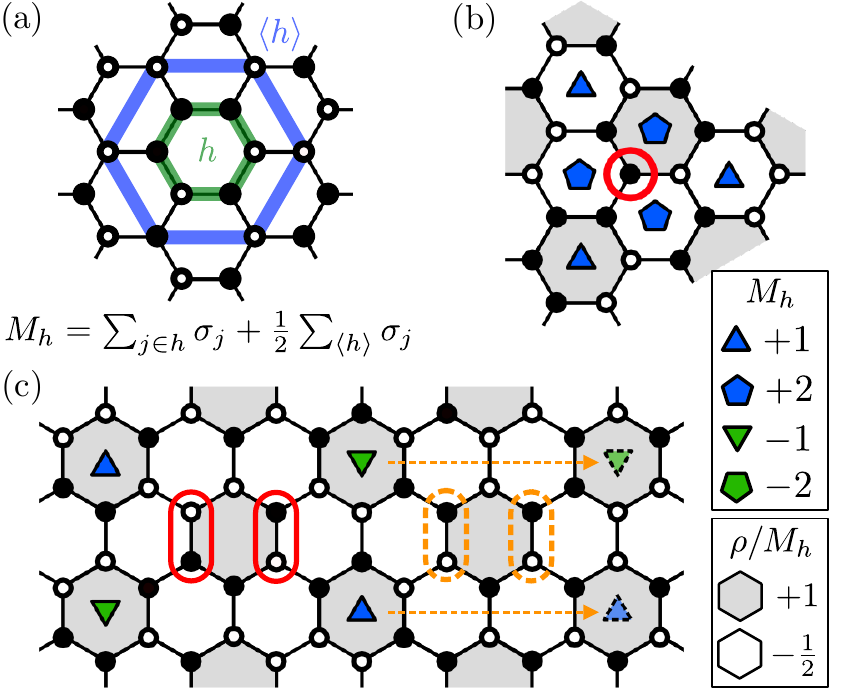}
\caption{Ground states of the honeycomb model fulfill the constraint $M_h=0$, illustrated in (a). A single spin flip, as indicated by a red circle in (b) preserves total charge $\rho$ as well as all dipole moments. The minimal-energy excitation, shown in (c) hence creates four distinct defects. These can be moved in pairs by flipping another four spins as indicated by orange dashed lines.}
\label{fig:model}
\end{figure}

Here, we present an Ising model exhibiting a fractonic spin liquid regime.
The model has an extensive degeneracy of ground states, which may be thought of as the classical limit of a gapless fracton phase.
This is distinct from fracton topological order, in which there is a subextensive degeneracy of locally indistinguishable quantum ground states.
The excitations of the model are fractons appearing at the corners of membranes of flipped spins. Pairs of fractons can be bound into lineons, which can move along a certain lattice direction, but not along the perpendicular direction.
We identify microscopic resonance processes which take the system between different ground states, consideration of which 
yields a lower bound on the observed entropy.
In order to simulate the model's properties 
we devise -- to our knowledge, the first --  cluster algorithm for Monte Carlo simulations of fractons, which is purpose-built for the study of fractonic Ising models.
This allows numerical access to relatively large systems and low temperatures. 

Simulating the model, we find that it exhibits a first order
phase transition at low temperatures, but that the low temperature state nevertheless lacks signs of conventional order.
Instead, the ground state correlations exhibit four-fold
pinch points in momentum space, which are characteristic
of systems described by a gauge theory of tensor fields \cite{prem2019pinch_points}.

\paragraph{Fractons in the honeycomb model.} 
We consider an Ising model on the honeycomb lattice with Hamiltonian
\begin{equation}
    H = \frac{J}{2}\sum_{h} M_h^2, ~~M_h=\sum_{j\in h} \sigma_j +\frac{1}{2}\sum_{j\in\expval{h}} \sigma_j,
    \label{eq:model}
\end{equation}
which is a sum over constraints $M_h$, defined on hexagons $h$ and their exteriors $\expval{h}$ as illustrated in \autoref{fig:model} (a). A Hamiltonian of this form was first considered for $O(3)$ Heisenberg spins in Ref.~\onlinecite{benton2021topo}. 

There, it was shown that the system upon coarse graining can be described in terms of a suitably defined rank-2 tensor field $m$, subjected to a Gauss law
\begin{equation}
    \partial_\mu \partial_\nu m^{\mu\nu} = \rho
\end{equation}
with ${\rm Tr}[m]=0$. The structure of this Gauss's law 
implies conservation of the dipole moments of charge density, and hence fractonic excitations, as we discusss further below.

The field $m$ relates to the microscopic degrees of freedom 
via a staggered, position dependent mapping.
As a result, the relationship between 
the charge $\rho$ and the microscopic constraint $M_h$   breaks lattice symmetry by hand. We choose a subset of hexagons such that each site is a member of exactly one of them. One possible choice is illustrated in \autoref{fig:model} by a darker shade of some hexagons. Denoting this subset as $+$ hexagons and the rest as $-$, the charge is then defined as
\begin{equation}
    \rho =
    \begin{cases}
        \phantom{-\frac{1}{2}} M_h & h \text{ is $+$ hexagon} \\
        -\frac{1}{2} M_h & \text{$h$ is $-$ hexagon}
    \end{cases}.
\end{equation}
Charges and moments of the higher-rank gauge theory can be determined explicitly  from the microscopic model. The single spin flip, shown in \autoref{fig:model} (b), preserves both total charge $\rho$ and the dipole moments $d^\nu = r^\nu \rho$. The lowest order moment of the charge distribution which changes is the quadrupole moment $q^{\mu\nu}=r^\mu r^\nu\rho$.  

An excitation with the lowest (non-zero) energy of \autoref{eq:model} involves four defects (hexagons with $M_h=\pm1$); it can be constructed from four spin flips, \autoref{fig:model} (c). These defects are fractons: no single defect can be moved by any local combination of spin flips since that would change the total dipole moment. Pairs of (oppositely charged) fractons are lineons since they can be moved in the direction perpendicular to their dipole moment by flipping another four spins on the next-next hexagon, as indicated in \autoref{fig:model} (c) in dashed-orange. 
Generally, fractons in our model appear at the corners of a ``membrane'' of flipped bonds.
This, along with the ability to bind fractons into 
lineons implies this is a ``Type-I'' fracton model,
constrasting with ``Type-II'' models where fractons appear
on fractal structures without the possibility of mobile bound 
states \cite{vijay2016}.

\begin{figure}
\includegraphics{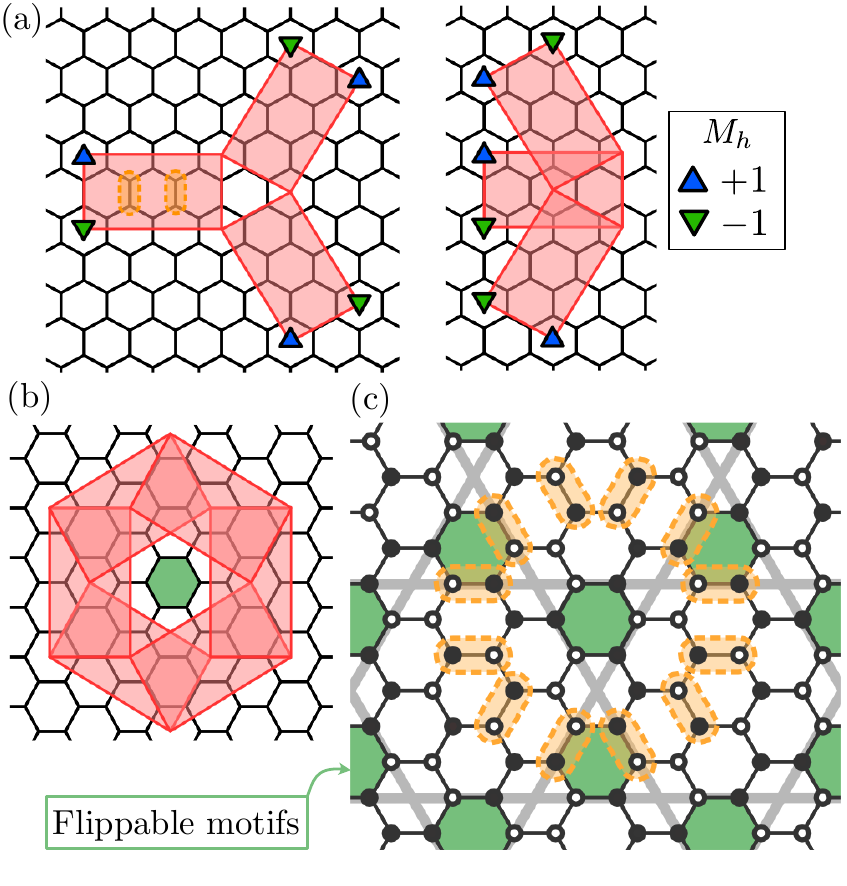}
\caption{
Splitting of lineons and 
local resonance moves.
(a) Pairs of fractons form lineons, mobile along a one-dimensional submanifold.
These lineons can split in two, 
either the forwards or backward directions.
(b) A local combination of 
six membranes, can create, split
and recombine lineons, in such a way
as to reach a new ground state
configuration.
This amounts to flipping 24 spins.
Application of this move to
the ground state in (c) does not create any excitation and could be centered around any of the green shaded hexagons. The move can be understood as a local combination of six membranes as shown in (b), where each corner overlaps with exactly one other corner of opposite charge.
This is possible because as shown in (c), a single two-fracton lineon can split into two both in the forward (left) and in the backward direction (right).}
\label{fig:move}
\end{figure}

\begin{figure*}
\includegraphics{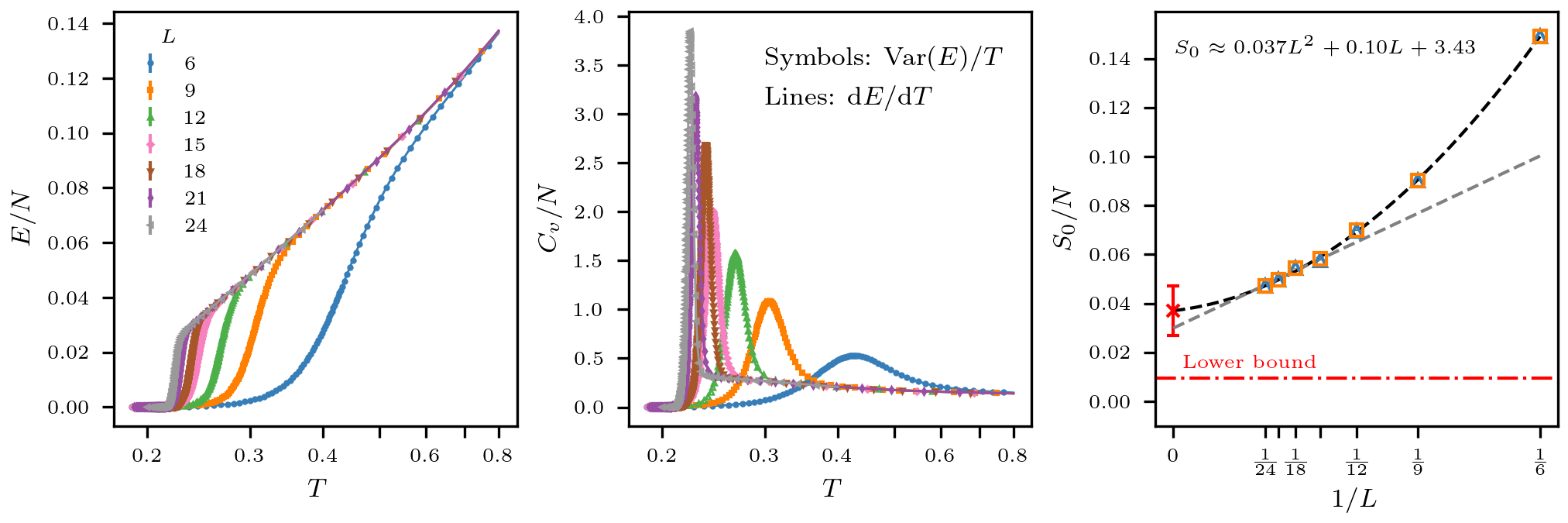}
\caption{
Thermodynamics of the fractonic Ising model.
Energy $E$, Specific heat $C_v$, and residual entropy $S_0$ are obtained from Monte-Carlo simulations. Energy/specific heat are consistent with a first-order transition from a high-temperature paramagnetic phase to a low-temperature phase where the constraint $M_h=0$ is satisfied.
Finite-size extrapolation of the ground state entropy per site tends to a value significantly
higher than the lower bound discussed in the main text, 
indicating a strongly fluctuating regime below the first order transition.
}
\label{fig:thermo}
\end{figure*}

While the mapping between the microscopic model and the coarse-grained field $m^{\mu\nu}$ in principle generalizes to the Ising model, it is an open question whether the system still realizes a higher-rank Coulomb phase or whether restricting the degrees of freedom to be discrete yields a set of ground states the average over which no longer corresponds to the  deconfined phase of the gauge theory. For the case of the `conventional' Coulomb liquid, cases are known where the hard-spin Ising and Heisenberg behaviors are (e.g.\ pyrochlore ~\cite{Isakov_2004_classpyro}), and are not (e.g.\ kagome ~\cite{syozi_kag,kano_kagome,Garanin_Canals,Chern_kagome}) , in accord with that of  the soft-spin theory.  In the following, we  argue for the former case, that is that the Ising model [\autoref{eq:model}] does also realize a higher-rank Coulomb regime. To this end, we first show explicitly that the number of ground states grows exponentially with the number of sites $N$ by identifying local ``resonance processses'' between different ground states. Second, using a novel cluster Monte-Carlo algorithm that moves pairs of defects we gain access to the thermodynamics of the model at large system size and low temperature. This allows us to both quantitatively extrapolate the residual entropy and compute the low-temperature correlations.

\paragraph{Extensive ground-state degeneracy.}
For periodic boundary conditions, the fact that pairs of fractons are lineons already implies a subextensive ground state entropy: we can create a pair of lineons, move one of them around the system in a nontrivial way and annihilate the pair again, reaching a different ground state. In addition to moving a lineon in a particular direction, we can also split it into two, as shown in \autoref{fig:move} (a). Note that such a move is only possible because of sixfold rotation symmetry and would not be possible with cubic symmetry and more generally in the absence of at least three distinct orientations for the membranes. Crucially, in our case it can be done in two ways which we call forward and backwards split, shown on the left and right of \autoref{fig:move}
 (c) respectively. 
The backwards split is particularly important since it allows us to close the worldline of lineons \emph{locally}, resulting in a local move between different ground states. The minimal such move is shown in \autoref{fig:move} (b) and is a combination of six membranes such that each corner defect is annihilated with exactly one other corner of opposite charge. This corresponds to flipping 24 spins simultaneously, as indicated in dashed-orange in panel (c) of the same figure. There, we also explicitly show a state with a finite density of such flippable motifs. It is constructed by starting form the Ne\'el state (which has zero energy) and flipping spins along lines as indicated in gray, which also preserves the ground state constraint. The 24-spin move as shown in \autoref{fig:move} (b,c) could be centered on any of the hexagons colored in green, and non-overlapping motifs can be flipped independently. This establishes a number of states exponential in the number of sites $N$ and in particular implies a lower bound on the residual entropy
\begin{equation}
    S_0 \geq \frac{N}{72} \log(2).
    \label{eq:s0-min}
\end{equation}
Since this is a particular construction and there is a wealth of possibilities to combine membranes locally such that they create no defects, we do not expect this bound to be tight. A more quantitative estimate can be obtained from Monte-Carlo simulations (cf. \autoref{fig:thermo}) as discussed in the following.

\begin{figure}
\includegraphics{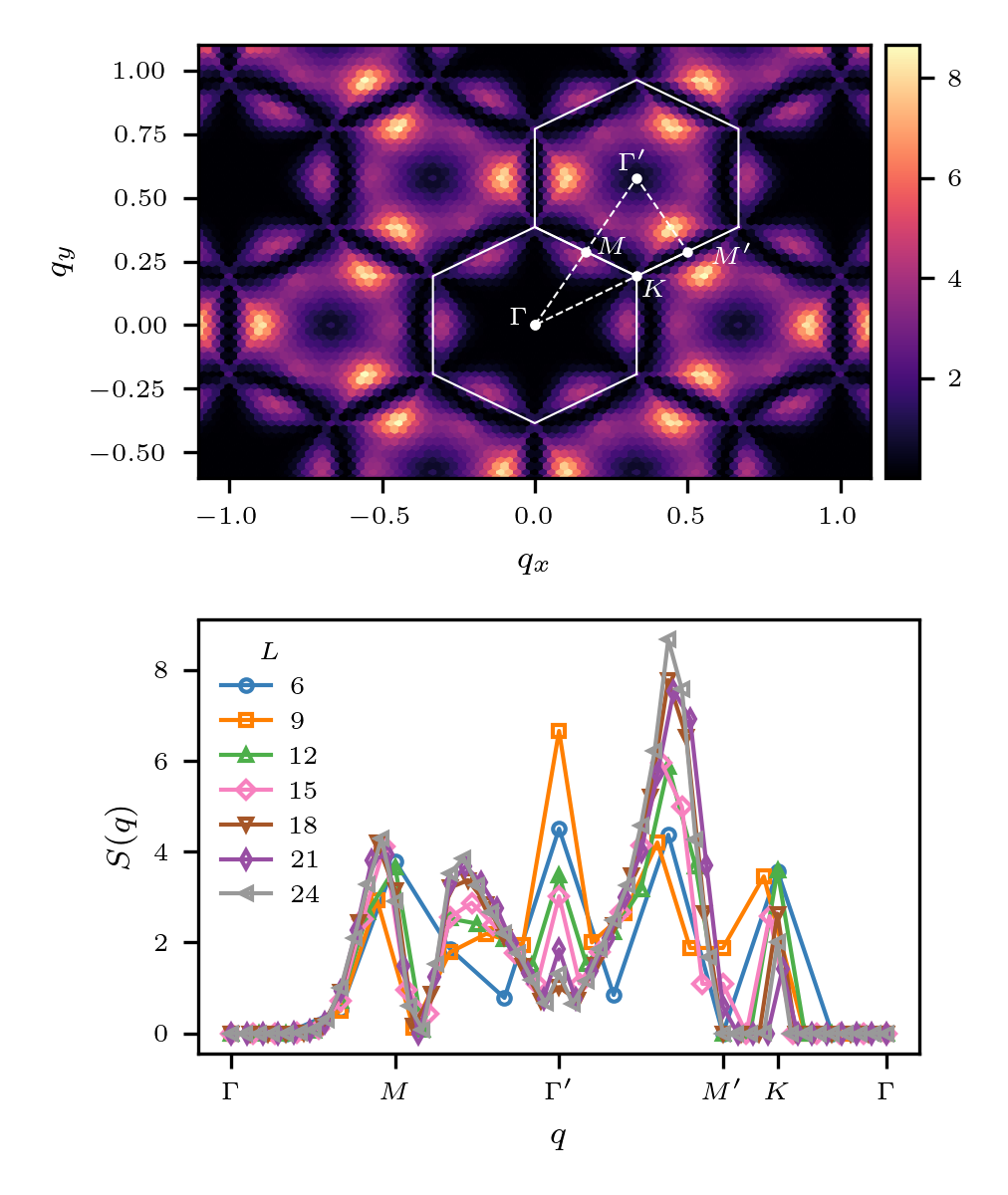}
\caption{The structure factor in the low-temperature phase at the system sizes studied shows no sharp features except clearly visible four-fold pinch points. These pinch points are the signature
 of an emergent higher-rank Gauss law \cite{prem2019pinch_points}
 and are of the same form as found for a higher-rank spin liquid in a related continuous spin model \cite{benton2021topo}. }
\label{fig:corr}
\end{figure}

\paragraph{Cluster Monte-Carlo algorithm.} 
It is well established that in the presence of fracton excitations, any local algorithm will have a rapidly diverging relaxation time at low tempearture \cite{newman1999glassy, chamon2005fractons, castelnovo2011glass}. Since single defects are immobile, local algorithms generally fail to anneal them out at low temperature and to gain access of the thermodynamics of \autoref{eq:model}, clearly a cluster algorithm is desirable. 
We have designed such an algorithm, which moves pairs of defects by effectively attempting to span one row of a membrane as shown in \autoref{fig:move} (c). In the crucial step, the algorithm starts by flipping two bonds on a single hexagon and accepting this move with metropolis probability. In the case that the move is \emph{rejected}, the algorithm attempts to flip another four spins [as indicated in dashed orange in \autoref{fig:move} (c)], again trying to accept the move with metropolis probability, multiplied by a factor to account for the probability of rejecting the first move. This additional factor is needed to ensure detailed balance \cite{krauthMonteCarlo}. This is repeated until either the cluster is accepted or is ultimately rejected while spanning the full (linear) system size. If the move is accepted with zero energy cost, it either moves a distant pair of defects by one step in the direction perpendicular to the long side of the (single-row) membrane or it changes the ground state sector. 
While the relaxation time of this algorithm still scales significantly with system size at low temperature (we estimate $\tau\sim L^{7.4}$), it constitutes a major improvement over local dynamics. 
To demonstrate this, we performed a simulated annealing simulation using 100 intermediate temperatures between $T=J$ and $T=J/10$ and $10^6$ sweeps per temperature. When using a spin-flip metropolis algorithm at $L=18$, out of 500 runs not a single one manages to anneal out all defects. In contrast, our cluster algorithm in the same configuration reaches zero energy in all cases. 
When augmented with the local 24 spin move shown in \autoref{fig:move} (b-c), accepted with metropolis probability, and using feedback-optimized parallel tempering \cite{katzgraber2006feedback}, we are able to equilibrate systems with up to $N=1152$ ($L=24$) spins in the ground state regime. To ensure equilibration, we compute the specific heat both directly from energy fluctuations and also as the derivative of internal energy with respect to temperature, and verify that these two estimators agree (cf. \autoref{fig:thermo}). This is considered a ``stringent criterion'' and used for example in computational studies of model glass formers \cite{berthier2023review}.

\paragraph{Thermodynamic properties and low-temperature correlations.} We use the setup described in the previous paragraph to study the thermodynamic properties of \autoref{eq:model} as a function of temperature for a range of system sizes as shown in \autoref{fig:thermo}. Both internal energy and specific heat are compatible with a first-order transition from a high-temperature paramagnetic phase to a low-temperature phase corresponding to the ground state regime $\expval{E}\approx 0$. Integrating the specific heat from high temperature yields an estimate of the residual entropy per site, $S_0/N$ for each system size. In the figure, we show the result of integrating both the variance of the internal energy (blue triangles) as well as its derivative with respect to temperature (orange squares). Fitting the system size dependence using both a linear fit to the last four sizes and a quadratic fit to all system sizes yields an estimate of $S_0/N=0.037\pm 0.010$. This is quite a bit above the lower bound derived above [\autoref{eq:s0-min}], but also comes with a large error bar due to significant finite size effects. The presence of these even at relatively large system sizes is not surprising given the size of the smallest local move as shown in \autoref{fig:move} (b,c). 
As a rough quantification of the importance of finite-size effects, we note that 
taking into account periodic boundary conditions, the largest system size $L=24$ has $\lfloor L/2\rfloor=12$ independent degrees of freedom on the boundary, compared to $\lfloor L/6 \rfloor^2\simeq 16$ in the bulk.

Finally, we show the structure factor within the low-temperature phase in \autoref{fig:corr}. It is fully consistent with a low-temperature higher-rank Coulomb phase and shows four-fold pinch-points at the zone boundaries \cite{prem2019pinch_points}, which are sharp (that is exactly one pixel) for all system sizes considered. 
% The structure factor is even quantitatively in excellent agreement even with the result from a soft-spin calculation \cite{benton2021topo}. 

The presence of a sharp transition as a function of temperature is somewhat surprising since in two dimensions, the higher-rank Coulomb liquid expected to describe the ground state regime is continuously connected to the paramagnet. 
It is however not inconsistent with a low-temperature liquid phase since a first-order transition is always possible also between continuously connected phases as demonstrated by the famous transition between gaseous and liquid water.
An alternative possibility is that the low-temperature phase is a so called ``fragmented liquid'' \cite{brooks14prx}, that is the set of ground states, although extensive, breaks some symmetry on average. A possible hint in this direction is that the maximum of the structure factor along the line cut shown in \autoref{fig:corr} scales roughly with linear system size. However, as discussed already above there are still significant finite size effects. Ultimately, it is impossible to exclude this possibility of fragmentation in the absence of a more efficient algorithm and we leave this question open for future studies.

\paragraph{Conclusion.}
In summary, we have demonstrated the
appearance of a fractonic spin liquid in the low temperature state
of an Ising model on the honeycomb lattice. 
This low temperature state is separated from the high temperature
paramagnet by a first order phase
transition, and exhibits correlations matching those predicted for a Coulomb phase of rank-2 electric fields with scalar charges \cite{prem2019pinch_points}.
Elementary excitations are Type-I fractons, appearing at the corners of membranes of flipped spins.

The discovery of a relatively simple Ising model, with finite-range, two-body interactions establishes a useful platform for the further exploration of fractonic physics. This could include the perturbative introduction of quantum effects via transverse fields or transverse exchange. 
This may be a better setting in which to study quantum effects on fractons than in the Heisenberg models suggested in Ref.~\onlinecite{benton2021topo}, for which numerical calculations suggest that quantum fluctuations wash out the 
multifold pinch points \cite{NiggemannPRL2023}.
Here, the emergent Gauss's law is protected by a finite gap, so it may be more robust.
Even if instanton effects drive the emergent gauge theory into a confined phase (as they do for the ordinary U(1) gauge theory in 2+1 D), the low temperature physics can still show interesting features related to the liquid phase.

Our purpose-built Monte Carlo algorithm provides a template for future numerical studies of Type-I fractonic models.
Future work could address in more detail the topics of relaxation and
disorder-free glassiness.
The successful demonstration of an Ising fracton spin liquid, based on a Hamiltonian originally constructed for continuous spins \cite{benton2021topo}, also raises the question of whether other classical spin liquids with higher-moment conservation laws \cite{yan_classification_short, yan_classification_long} have Ising realizations, and what their properties may be. 

Given the rapid progress in physical simulation platforms based on Rydberg atoms and superconducting qubits, the dynamics of fractons in our model may not be far from exploration in the laboratory.\\

\begin{acknowledgments}
	This work was in part supported by the Deutsche Forschungsgemeinschaft under grants  SFB 1143 (project-id 247310070) and the cluster of excellence 
ct.qmat (EXC 2147, project-id 390858490).
\end{acknowledgments}

\bibliography{fractons.bib}

\begin{thebibliography}{41}%
\makeatletter
\providecommand \@ifxundefined [1]{%
 \@ifx{#1\undefined}
}%
\providecommand \@ifnum [1]{%
 \ifnum #1\expandafter \@firstoftwo
 \else \expandafter \@secondoftwo
 \fi
}%
\providecommand \@ifx [1]{%
 \ifx #1\expandafter \@firstoftwo
 \else \expandafter \@secondoftwo
 \fi
}%
\providecommand \natexlab [1]{#1}%
\providecommand \enquote  [1]{``#1''}%
\providecommand \bibnamefont  [1]{#1}%
\providecommand \bibfnamefont [1]{#1}%
\providecommand \citenamefont [1]{#1}%
\providecommand \href@noop [0]{\@secondoftwo}%
\providecommand \href [0]{\begingroup \@sanitize@url \@href}%
\providecommand \@href[1]{\@@startlink{#1}\@@href}%
\providecommand \@@href[1]{\endgroup#1\@@endlink}%
\providecommand \@sanitize@url [0]{\catcode `\\12\catcode `\$12\catcode
  `\&12\catcode `\#12\catcode `\^12\catcode `\_12\catcode `\%12\relax}%
\providecommand \@@startlink[1]{}%
\providecommand \@@endlink[0]{}%
\providecommand \url  [0]{\begingroup\@sanitize@url \@url }%
\providecommand \@url [1]{\endgroup\@href {#1}{\urlprefix }}%
\providecommand \urlprefix  [0]{URL }%
\providecommand \Eprint [0]{\href }%
\providecommand \doibase [0]{http://dx.doi.org/}%
\providecommand \selectlanguage [0]{\@gobble}%
\providecommand \bibinfo  [0]{\@secondoftwo}%
\providecommand \bibfield  [0]{\@secondoftwo}%
\providecommand \translation [1]{[#1]}%
\providecommand \BibitemOpen [0]{}%
\providecommand \bibitemStop [0]{}%
\providecommand \bibitemNoStop [0]{.\EOS\space}%
\providecommand \EOS [0]{\spacefactor3000\relax}%
\providecommand \BibitemShut  [1]{\csname bibitem#1\endcsname}%
\let\auto@bib@innerbib\@empty
%</preamble>
\bibitem [{\citenamefont {Newman}\ and\ \citenamefont
  {Moore}(1999)}]{newman1999glassy}%
  \BibitemOpen
  \bibfield  {author} {\bibinfo {author} {\bibfnamefont {M.~E.~J.}\
  \bibnamefont {Newman}}\ and\ \bibinfo {author} {\bibfnamefont
  {C.}~\bibnamefont {Moore}},\ }\href {\doibase 10.1103/PhysRevE.60.5068}
  {\bibfield  {journal} {\bibinfo  {journal} {Phys. Rev. E}\ }\textbf {\bibinfo
  {volume} {60}},\ \bibinfo {pages} {5068} (\bibinfo {year}
  {1999})}\BibitemShut {NoStop}%
\bibitem [{\citenamefont {Chamon}(2005)}]{chamon2005fractons}%
  \BibitemOpen
  \bibfield  {author} {\bibinfo {author} {\bibfnamefont {C.}~\bibnamefont
  {Chamon}},\ }\href {\doibase 10.1103/PhysRevLett.94.040402} {\bibfield
  {journal} {\bibinfo  {journal} {Phys. Rev. Lett.}\ }\textbf {\bibinfo
  {volume} {94}},\ \bibinfo {pages} {040402} (\bibinfo {year}
  {2005})}\BibitemShut {NoStop}%
\bibitem [{\citenamefont {Bravyi}\ \emph {et~al.}(2011)\citenamefont {Bravyi},
  \citenamefont {Leemhuis},\ and\ \citenamefont {Terhal}}]{bravyi2011}%
  \BibitemOpen
  \bibfield  {author} {\bibinfo {author} {\bibfnamefont {S.}~\bibnamefont
  {Bravyi}}, \bibinfo {author} {\bibfnamefont {B.}~\bibnamefont {Leemhuis}}, \
  and\ \bibinfo {author} {\bibfnamefont {B.~M.}\ \bibnamefont {Terhal}},\
  }\href {\doibase https://doi.org/10.1016/j.aop.2010.11.002} {\bibfield
  {journal} {\bibinfo  {journal} {Annals of Physics}\ }\textbf {\bibinfo
  {volume} {326}},\ \bibinfo {pages} {839} (\bibinfo {year}
  {2011})}\BibitemShut {NoStop}%
\bibitem [{\citenamefont {Bravyi}\ and\ \citenamefont
  {Haah}(2013)}]{bravyi2013}%
  \BibitemOpen
  \bibfield  {author} {\bibinfo {author} {\bibfnamefont {S.}~\bibnamefont
  {Bravyi}}\ and\ \bibinfo {author} {\bibfnamefont {J.}~\bibnamefont {Haah}},\
  }\href {\doibase 10.1103/PhysRevLett.111.200501} {\bibfield  {journal}
  {\bibinfo  {journal} {Phys. Rev. Lett.}\ }\textbf {\bibinfo {volume} {111}},\
  \bibinfo {pages} {200501} (\bibinfo {year} {2013})}\BibitemShut {NoStop}%
\bibitem [{\citenamefont {Castelnovo}\ and\ \citenamefont
  {Chamon}(2012)}]{castelnovo2011glass}%
  \BibitemOpen
  \bibfield  {author} {\bibinfo {author} {\bibfnamefont {C.}~\bibnamefont
  {Castelnovo}}\ and\ \bibinfo {author} {\bibfnamefont {C.}~\bibnamefont
  {Chamon}},\ }\href {\doibase 10.1080/14786435.2011.609152} {\bibfield
  {journal} {\bibinfo  {journal} {Philosophical Magazine}\ }\textbf {\bibinfo
  {volume} {92}},\ \bibinfo {pages} {304} (\bibinfo {year} {2012})},\ \Eprint
  {http://arxiv.org/abs/https://doi.org/10.1080/14786435.2011.609152}
  {https://doi.org/10.1080/14786435.2011.609152} \BibitemShut {NoStop}%
\bibitem [{\citenamefont {Yoshida}(2013)}]{yoshida2013}%
  \BibitemOpen
  \bibfield  {author} {\bibinfo {author} {\bibfnamefont {B.}~\bibnamefont
  {Yoshida}},\ }\href {\doibase 10.1103/PhysRevB.88.125122} {\bibfield
  {journal} {\bibinfo  {journal} {Phys. Rev. B}\ }\textbf {\bibinfo {volume}
  {88}},\ \bibinfo {pages} {125122} (\bibinfo {year} {2013})}\BibitemShut
  {NoStop}%
\bibitem [{\citenamefont {Vijay}\ \emph {et~al.}(2015)\citenamefont {Vijay},
  \citenamefont {Haah},\ and\ \citenamefont {Fu}}]{vijay2015}%
  \BibitemOpen
  \bibfield  {author} {\bibinfo {author} {\bibfnamefont {S.}~\bibnamefont
  {Vijay}}, \bibinfo {author} {\bibfnamefont {J.}~\bibnamefont {Haah}}, \ and\
  \bibinfo {author} {\bibfnamefont {L.}~\bibnamefont {Fu}},\ }\href {\doibase
  10.1103/PhysRevB.92.235136} {\bibfield  {journal} {\bibinfo  {journal} {Phys.
  Rev. B}\ }\textbf {\bibinfo {volume} {92}},\ \bibinfo {pages} {235136}
  (\bibinfo {year} {2015})}\BibitemShut {NoStop}%
\bibitem [{\citenamefont {Vijay}\ \emph {et~al.}(2016)\citenamefont {Vijay},
  \citenamefont {Haah},\ and\ \citenamefont {Fu}}]{vijay2016}%
  \BibitemOpen
  \bibfield  {author} {\bibinfo {author} {\bibfnamefont {S.}~\bibnamefont
  {Vijay}}, \bibinfo {author} {\bibfnamefont {J.}~\bibnamefont {Haah}}, \ and\
  \bibinfo {author} {\bibfnamefont {L.}~\bibnamefont {Fu}},\ }\href {\doibase
  10.1103/PhysRevB.94.235157} {\bibfield  {journal} {\bibinfo  {journal} {Phys.
  Rev. B}\ }\textbf {\bibinfo {volume} {94}},\ \bibinfo {pages} {235157}
  (\bibinfo {year} {2016})}\BibitemShut {NoStop}%
\bibitem [{\citenamefont {Haah}(2011)}]{haah2011code}%
  \BibitemOpen
  \bibfield  {author} {\bibinfo {author} {\bibfnamefont {J.}~\bibnamefont
  {Haah}},\ }\href {\doibase 10.1103/PhysRevA.83.042330} {\bibfield  {journal}
  {\bibinfo  {journal} {Phys. Rev. A}\ }\textbf {\bibinfo {volume} {83}},\
  \bibinfo {pages} {042330} (\bibinfo {year} {2011})}\BibitemShut {NoStop}%
\bibitem [{\citenamefont {Pretko}\ \emph {et~al.}(2020)\citenamefont {Pretko},
  \citenamefont {Chen},\ and\ \citenamefont {You}}]{pretko2020review}%
  \BibitemOpen
  \bibfield  {author} {\bibinfo {author} {\bibfnamefont {M.}~\bibnamefont
  {Pretko}}, \bibinfo {author} {\bibfnamefont {X.}~\bibnamefont {Chen}}, \ and\
  \bibinfo {author} {\bibfnamefont {Y.}~\bibnamefont {You}},\ }\href {\doibase
  10.1142/S0217751X20300033} {\bibfield  {journal} {\bibinfo  {journal}
  {International Journal of Modern Physics A}\ }\textbf {\bibinfo {volume}
  {35}},\ \bibinfo {pages} {2030003} (\bibinfo {year} {2020})},\ \Eprint
  {http://arxiv.org/abs/https://doi.org/10.1142/S0217751X20300033}
  {https://doi.org/10.1142/S0217751X20300033} \BibitemShut {NoStop}%
\bibitem [{\citenamefont {Nandkishore}\ and\ \citenamefont
  {Hermele}(2019)}]{nandkoshore2019review}%
  \BibitemOpen
  \bibfield  {author} {\bibinfo {author} {\bibfnamefont {R.~M.}\ \bibnamefont
  {Nandkishore}}\ and\ \bibinfo {author} {\bibfnamefont {M.}~\bibnamefont
  {Hermele}},\ }\href {\doibase 10.1146/annurev-conmatphys-031218-013604}
  {\bibfield  {journal} {\bibinfo  {journal} {Annual Review of Condensed Matter
  Physics}\ }\textbf {\bibinfo {volume} {10}},\ \bibinfo {pages} {295}
  (\bibinfo {year} {2019})},\ \Eprint
  {http://arxiv.org/abs/https://doi.org/10.1146/annurev-conmatphys-031218-013604}
  {https://doi.org/10.1146/annurev-conmatphys-031218-013604} \BibitemShut
  {NoStop}%
\bibitem [{\citenamefont {Xu}(2006)}]{xu2006hrgt}%
  \BibitemOpen
  \bibfield  {author} {\bibinfo {author} {\bibfnamefont {C.}~\bibnamefont
  {Xu}},\ }\href {\doibase 10.1103/PhysRevB.74.224433} {\bibfield  {journal}
  {\bibinfo  {journal} {Phys. Rev. B}\ }\textbf {\bibinfo {volume} {74}},\
  \bibinfo {pages} {224433} (\bibinfo {year} {2006})}\BibitemShut {NoStop}%
\bibitem [{\citenamefont {Xu}\ and\ \citenamefont {Ho\ifmmode~\check{r}\else
  \v{r}\fi{}ava}(2010)}]{xu2010prd}%
  \BibitemOpen
  \bibfield  {author} {\bibinfo {author} {\bibfnamefont {C.}~\bibnamefont
  {Xu}}\ and\ \bibinfo {author} {\bibfnamefont {P.}~\bibnamefont
  {Ho\ifmmode~\check{r}\else \v{r}\fi{}ava}},\ }\href {\doibase
  10.1103/PhysRevD.81.104033} {\bibfield  {journal} {\bibinfo  {journal} {Phys.
  Rev. D}\ }\textbf {\bibinfo {volume} {81}},\ \bibinfo {pages} {104033}
  (\bibinfo {year} {2010})}\BibitemShut {NoStop}%
\bibitem [{\citenamefont {Pretko}(2017{\natexlab{a}})}]{pretko2017}%
  \BibitemOpen
  \bibfield  {author} {\bibinfo {author} {\bibfnamefont {M.}~\bibnamefont
  {Pretko}},\ }\href {\doibase 10.1103/PhysRevB.95.115139} {\bibfield
  {journal} {\bibinfo  {journal} {Phys. Rev. B}\ }\textbf {\bibinfo {volume}
  {95}},\ \bibinfo {pages} {115139} (\bibinfo {year}
  {2017}{\natexlab{a}})}\BibitemShut {NoStop}%
\bibitem [{\citenamefont {Pretko}(2017{\natexlab{b}})}]{pretko2017b}%
  \BibitemOpen
  \bibfield  {author} {\bibinfo {author} {\bibfnamefont {M.}~\bibnamefont
  {Pretko}},\ }\href {\doibase 10.1103/PhysRevB.96.035119} {\bibfield
  {journal} {\bibinfo  {journal} {Phys. Rev. B}\ }\textbf {\bibinfo {volume}
  {96}},\ \bibinfo {pages} {035119} (\bibinfo {year}
  {2017}{\natexlab{b}})}\BibitemShut {NoStop}%
\bibitem [{\citenamefont {Pretko}(2017{\natexlab{c}})}]{Pretko2017gravity}%
  \BibitemOpen
  \bibfield  {author} {\bibinfo {author} {\bibfnamefont {M.}~\bibnamefont
  {Pretko}},\ }\href {\doibase 10.1103/PhysRevD.96.024051} {\bibfield
  {journal} {\bibinfo  {journal} {Phys. Rev. D}\ }\textbf {\bibinfo {volume}
  {96}},\ \bibinfo {pages} {024051} (\bibinfo {year}
  {2017}{\natexlab{c}})}\BibitemShut {NoStop}%
\bibitem [{\citenamefont {Pretko}\ and\ \citenamefont
  {Radzihovsky}(2018)}]{Pretko2018elasticity}%
  \BibitemOpen
  \bibfield  {author} {\bibinfo {author} {\bibfnamefont {M.}~\bibnamefont
  {Pretko}}\ and\ \bibinfo {author} {\bibfnamefont {L.}~\bibnamefont
  {Radzihovsky}},\ }\href {\doibase 10.1103/PhysRevLett.120.195301} {\bibfield
  {journal} {\bibinfo  {journal} {Phys. Rev. Lett.}\ }\textbf {\bibinfo
  {volume} {120}},\ \bibinfo {pages} {195301} (\bibinfo {year}
  {2018})}\BibitemShut {NoStop}%
\bibitem [{\citenamefont {Sous}\ and\ \citenamefont
  {Pretko}(2020)}]{sous2020holes}%
  \BibitemOpen
  \bibfield  {author} {\bibinfo {author} {\bibfnamefont {J.}~\bibnamefont
  {Sous}}\ and\ \bibinfo {author} {\bibfnamefont {M.}~\bibnamefont {Pretko}},\
  }\href {\doibase 10.1038/s41535-020-00278-2} {\bibfield  {journal} {\bibinfo
  {journal} {npj Quantum Materials}\ }\textbf {\bibinfo {volume} {5}},\
  \bibinfo {pages} {81} (\bibinfo {year} {2020})}\BibitemShut {NoStop}%
\bibitem [{\citenamefont {You}\ \emph {et~al.}(2020)\citenamefont {You},
  \citenamefont {Bi},\ and\ \citenamefont {Pretko}}]{You2020plaquette}%
  \BibitemOpen
  \bibfield  {author} {\bibinfo {author} {\bibfnamefont {Y.}~\bibnamefont
  {You}}, \bibinfo {author} {\bibfnamefont {Z.}~\bibnamefont {Bi}}, \ and\
  \bibinfo {author} {\bibfnamefont {M.}~\bibnamefont {Pretko}},\ }\href
  {\doibase 10.1103/PhysRevResearch.2.013162} {\bibfield  {journal} {\bibinfo
  {journal} {Phys. Rev. Res.}\ }\textbf {\bibinfo {volume} {2}},\ \bibinfo
  {pages} {013162} (\bibinfo {year} {2020})}\BibitemShut {NoStop}%
\bibitem [{\citenamefont {Pretko}(2019)}]{Pretko2019circuits}%
  \BibitemOpen
  \bibfield  {author} {\bibinfo {author} {\bibfnamefont {M.}~\bibnamefont
  {Pretko}},\ }\href {\doibase 10.1103/PhysRevB.100.245103} {\bibfield
  {journal} {\bibinfo  {journal} {Phys. Rev. B}\ }\textbf {\bibinfo {volume}
  {100}},\ \bibinfo {pages} {245103} (\bibinfo {year} {2019})}\BibitemShut
  {NoStop}%
\bibitem [{\citenamefont {Hering}\ \emph {et~al.}(2021)\citenamefont {Hering},
  \citenamefont {Yan},\ and\ \citenamefont {Reuther}}]{Hering2021}%
  \BibitemOpen
  \bibfield  {author} {\bibinfo {author} {\bibfnamefont {M.}~\bibnamefont
  {Hering}}, \bibinfo {author} {\bibfnamefont {H.}~\bibnamefont {Yan}}, \ and\
  \bibinfo {author} {\bibfnamefont {J.}~\bibnamefont {Reuther}},\ }\href
  {\doibase 10.1103/PhysRevB.104.064406} {\bibfield  {journal} {\bibinfo
  {journal} {Phys. Rev. B}\ }\textbf {\bibinfo {volume} {104}},\ \bibinfo
  {pages} {064406} (\bibinfo {year} {2021})}\BibitemShut {NoStop}%
\bibitem [{\citenamefont {Yan}\ and\ \citenamefont {Reuther}(2022)}]{Yan2022}%
  \BibitemOpen
  \bibfield  {author} {\bibinfo {author} {\bibfnamefont {H.}~\bibnamefont
  {Yan}}\ and\ \bibinfo {author} {\bibfnamefont {J.}~\bibnamefont {Reuther}},\
  }\href {\doibase 10.1103/PhysRevResearch.4.023175} {\bibfield  {journal}
  {\bibinfo  {journal} {Phys. Rev. Res.}\ }\textbf {\bibinfo {volume} {4}},\
  \bibinfo {pages} {023175} (\bibinfo {year} {2022})}\BibitemShut {NoStop}%
\bibitem [{\citenamefont {Myerson-Jain}\ \emph {et~al.}(2022)\citenamefont
  {Myerson-Jain}, \citenamefont {Yan}, \citenamefont {Weld},\ and\
  \citenamefont {Xu}}]{myerson2022rydberg}%
  \BibitemOpen
  \bibfield  {author} {\bibinfo {author} {\bibfnamefont {N.~E.}\ \bibnamefont
  {Myerson-Jain}}, \bibinfo {author} {\bibfnamefont {S.}~\bibnamefont {Yan}},
  \bibinfo {author} {\bibfnamefont {D.}~\bibnamefont {Weld}}, \ and\ \bibinfo
  {author} {\bibfnamefont {C.}~\bibnamefont {Xu}},\ }\href {\doibase
  10.1103/PhysRevLett.128.017601} {\bibfield  {journal} {\bibinfo  {journal}
  {Phys. Rev. Lett.}\ }\textbf {\bibinfo {volume} {128}},\ \bibinfo {pages}
  {017601} (\bibinfo {year} {2022})}\BibitemShut {NoStop}%
\bibitem [{\citenamefont {Giergiel}\ \emph {et~al.}(2022)\citenamefont
  {Giergiel}, \citenamefont {Lier}, \citenamefont {Sur\'owka},\ and\
  \citenamefont {Kosior}}]{giergiel2022bosehubbard}%
  \BibitemOpen
  \bibfield  {author} {\bibinfo {author} {\bibfnamefont {K.}~\bibnamefont
  {Giergiel}}, \bibinfo {author} {\bibfnamefont {R.}~\bibnamefont {Lier}},
  \bibinfo {author} {\bibfnamefont {P.}~\bibnamefont {Sur\'owka}}, \ and\
  \bibinfo {author} {\bibfnamefont {A.}~\bibnamefont {Kosior}},\ }\href
  {\doibase 10.1103/PhysRevResearch.4.023151} {\bibfield  {journal} {\bibinfo
  {journal} {Phys. Rev. Res.}\ }\textbf {\bibinfo {volume} {4}},\ \bibinfo
  {pages} {023151} (\bibinfo {year} {2022})}\BibitemShut {NoStop}%
\bibitem [{\citenamefont {Han}\ \emph {et~al.}(2022)\citenamefont {Han},
  \citenamefont {Patri},\ and\ \citenamefont {Kim}}]{han2022breathing}%
  \BibitemOpen
  \bibfield  {author} {\bibinfo {author} {\bibfnamefont {S.}~\bibnamefont
  {Han}}, \bibinfo {author} {\bibfnamefont {A.~S.}\ \bibnamefont {Patri}}, \
  and\ \bibinfo {author} {\bibfnamefont {Y.~B.}\ \bibnamefont {Kim}},\ }\href
  {\doibase 10.1103/PhysRevB.105.235120} {\bibfield  {journal} {\bibinfo
  {journal} {Phys. Rev. B}\ }\textbf {\bibinfo {volume} {105}},\ \bibinfo
  {pages} {235120} (\bibinfo {year} {2022})}\BibitemShut {NoStop}%
\bibitem [{\citenamefont {Yan}\ \emph {et~al.}(2020)\citenamefont {Yan},
  \citenamefont {Benton}, \citenamefont {Jaubert},\ and\ \citenamefont
  {Shannon}}]{yan2020}%
  \BibitemOpen
  \bibfield  {author} {\bibinfo {author} {\bibfnamefont {H.}~\bibnamefont
  {Yan}}, \bibinfo {author} {\bibfnamefont {O.}~\bibnamefont {Benton}},
  \bibinfo {author} {\bibfnamefont {L.~D.~C.}\ \bibnamefont {Jaubert}}, \ and\
  \bibinfo {author} {\bibfnamefont {N.}~\bibnamefont {Shannon}},\ }\href
  {\doibase 10.1103/PhysRevLett.124.127203} {\bibfield  {journal} {\bibinfo
  {journal} {Phys. Rev. Lett.}\ }\textbf {\bibinfo {volume} {124}},\ \bibinfo
  {pages} {127203} (\bibinfo {year} {2020})}\BibitemShut {NoStop}%
\bibitem [{\citenamefont {Benton}\ and\ \citenamefont
  {Moessner}(2021)}]{benton2021topo}%
  \BibitemOpen
  \bibfield  {author} {\bibinfo {author} {\bibfnamefont {O.}~\bibnamefont
  {Benton}}\ and\ \bibinfo {author} {\bibfnamefont {R.}~\bibnamefont
  {Moessner}},\ }\href {\doibase 10.1103/PhysRevLett.127.107202} {\bibfield
  {journal} {\bibinfo  {journal} {Phys. Rev. Lett.}\ }\textbf {\bibinfo
  {volume} {127}},\ \bibinfo {pages} {107202} (\bibinfo {year}
  {2021})}\BibitemShut {NoStop}%
\bibitem [{\citenamefont {Zhang}\ \emph {et~al.}(2022)\citenamefont {Zhang},
  \citenamefont {Buessen},\ and\ \citenamefont {Kim}}]{zhang2022dynamical}%
  \BibitemOpen
  \bibfield  {author} {\bibinfo {author} {\bibfnamefont {E.~Z.}\ \bibnamefont
  {Zhang}}, \bibinfo {author} {\bibfnamefont {F.~L.}\ \bibnamefont {Buessen}},
  \ and\ \bibinfo {author} {\bibfnamefont {Y.~B.}\ \bibnamefont {Kim}},\ }\href
  {\doibase 10.1103/PhysRevB.105.L060408} {\bibfield  {journal} {\bibinfo
  {journal} {Phys. Rev. B}\ }\textbf {\bibinfo {volume} {105}},\ \bibinfo
  {pages} {L060408} (\bibinfo {year} {2022})}\BibitemShut {NoStop}%
\bibitem [{\citenamefont {Prem}\ \emph {et~al.}(2018)\citenamefont {Prem},
  \citenamefont {Vijay}, \citenamefont {Chou}, \citenamefont {Pretko},\ and\
  \citenamefont {Nandkishore}}]{prem2019pinch_points}%
  \BibitemOpen
  \bibfield  {author} {\bibinfo {author} {\bibfnamefont {A.}~\bibnamefont
  {Prem}}, \bibinfo {author} {\bibfnamefont {S.}~\bibnamefont {Vijay}},
  \bibinfo {author} {\bibfnamefont {Y.-Z.}\ \bibnamefont {Chou}}, \bibinfo
  {author} {\bibfnamefont {M.}~\bibnamefont {Pretko}}, \ and\ \bibinfo {author}
  {\bibfnamefont {R.~M.}\ \bibnamefont {Nandkishore}},\ }\href {\doibase
  10.1103/PhysRevB.98.165140} {\bibfield  {journal} {\bibinfo  {journal} {Phys.
  Rev. B}\ }\textbf {\bibinfo {volume} {98}},\ \bibinfo {pages} {165140}
  (\bibinfo {year} {2018})}\BibitemShut {NoStop}%
\bibitem [{\citenamefont {{Isakov}}\ \emph {et~al.}(2004)\citenamefont
  {{Isakov}}, \citenamefont {{Gregor}}, \citenamefont {{Moessner}},\ and\
  \citenamefont {{Sondhi}}}]{Isakov_2004_classpyro}%
  \BibitemOpen
  \bibfield  {author} {\bibinfo {author} {\bibfnamefont {S.~V.}\ \bibnamefont
  {{Isakov}}}, \bibinfo {author} {\bibfnamefont {K.}~\bibnamefont {{Gregor}}},
  \bibinfo {author} {\bibfnamefont {R.}~\bibnamefont {{Moessner}}}, \ and\
  \bibinfo {author} {\bibfnamefont {S.~L.}\ \bibnamefont {{Sondhi}}},\ }\href
  {\doibase 10.1103/PhysRevLett.93.167204} {\bibfield  {journal} {\bibinfo
  {journal} {\prl}\ }\textbf {\bibinfo {volume} {93}},\ \bibinfo {eid} {167204}
  (\bibinfo {year} {2004})},\ \Eprint {http://arxiv.org/abs/cond-mat/0407004}
  {arXiv:cond-mat/0407004 [cond-mat.dis-nn]} \BibitemShut {NoStop}%
\bibitem [{\citenamefont {Sy\^ozi}(1951)}]{syozi_kag}%
  \BibitemOpen
  \bibfield  {author} {\bibinfo {author} {\bibfnamefont {I.}~\bibnamefont
  {Sy\^ozi}},\ }\href {\doibase 10.1143/ptp/6.3.306} {\bibfield  {journal}
  {\bibinfo  {journal} {Progress of Theoretical Physics}\ }\textbf {\bibinfo
  {volume} {6}},\ \bibinfo {pages} {306} (\bibinfo {year} {1951})},\ \Eprint
  {http://arxiv.org/abs/https://academic.oup.com/ptp/article-pdf/6/3/306/5239621/6-3-306.pdf}
  {https://academic.oup.com/ptp/article-pdf/6/3/306/5239621/6-3-306.pdf}
  \BibitemShut {NoStop}%
\bibitem [{\citenamefont {{Kan{\^o}}}\ and\ \citenamefont
  {{Naya}}(1953)}]{kano_kagome}%
  \BibitemOpen
  \bibfield  {author} {\bibinfo {author} {\bibfnamefont {K.}~\bibnamefont
  {{Kan{\^o}}}}\ and\ \bibinfo {author} {\bibfnamefont {S.}~\bibnamefont
  {{Naya}}},\ }\href {\doibase 10.1143/ptp/10.2.158} {\bibfield  {journal}
  {\bibinfo  {journal} {Progress of Theoretical Physics}\ }\textbf {\bibinfo
  {volume} {10}},\ \bibinfo {pages} {158} (\bibinfo {year} {1953})}\BibitemShut
  {NoStop}%
\bibitem [{\citenamefont {{Garanin}}\ and\ \citenamefont
  {{Canals}}(1999)}]{Garanin_Canals}%
  \BibitemOpen
  \bibfield  {author} {\bibinfo {author} {\bibfnamefont {D.~A.}\ \bibnamefont
  {{Garanin}}}\ and\ \bibinfo {author} {\bibfnamefont {B.}~\bibnamefont
  {{Canals}}},\ }\href {\doibase 10.1103/PhysRevB.59.443} {\bibfield  {journal}
  {\bibinfo  {journal} {\prb}\ }\textbf {\bibinfo {volume} {59}},\ \bibinfo
  {pages} {443} (\bibinfo {year} {1999})},\ \Eprint
  {http://arxiv.org/abs/cond-mat/9805362} {arXiv:cond-mat/9805362
  [cond-mat.stat-mech]} \BibitemShut {NoStop}%
\bibitem [{\citenamefont {{Chern}}\ and\ \citenamefont
  {{Moessner}}(2013)}]{Chern_kagome}%
  \BibitemOpen
  \bibfield  {author} {\bibinfo {author} {\bibfnamefont {G.-W.}\ \bibnamefont
  {{Chern}}}\ and\ \bibinfo {author} {\bibfnamefont {R.}~\bibnamefont
  {{Moessner}}},\ }\href {\doibase 10.1103/PhysRevLett.110.077201} {\bibfield
  {journal} {\bibinfo  {journal} {\prl}\ }\textbf {\bibinfo {volume} {110}},\
  \bibinfo {eid} {077201} (\bibinfo {year} {2013})},\ \Eprint
  {http://arxiv.org/abs/1207.4752} {arXiv:1207.4752 [cond-mat.str-el]}
  \BibitemShut {NoStop}%
\bibitem [{\citenamefont {Krauth}(1996)}]{krauthMonteCarlo}%
  \BibitemOpen
  \bibfield  {author} {\bibinfo {author} {\bibfnamefont {W.}~\bibnamefont
  {Krauth}}\ }(\bibinfo {year} {1996})\ \Eprint
  {http://arxiv.org/abs/arXiv:cond-mat/9612186} {arXiv:cond-mat/9612186}
  \BibitemShut {NoStop}%
\bibitem [{\citenamefont {Katzgraber}\ \emph {et~al.}(2006)\citenamefont
  {Katzgraber}, \citenamefont {Trebst}, \citenamefont {Huse},\ and\
  \citenamefont {Troyer}}]{katzgraber2006feedback}%
  \BibitemOpen
  \bibfield  {author} {\bibinfo {author} {\bibfnamefont {H.~G.}\ \bibnamefont
  {Katzgraber}}, \bibinfo {author} {\bibfnamefont {S.}~\bibnamefont {Trebst}},
  \bibinfo {author} {\bibfnamefont {D.~A.}\ \bibnamefont {Huse}}, \ and\
  \bibinfo {author} {\bibfnamefont {M.}~\bibnamefont {Troyer}},\ }\href
  {\doibase 10.1088/1742-5468/2006/03/P03018} {\bibfield  {journal} {\bibinfo
  {journal} {Journal of Statistical Mechanics: Theory and Experiment}\ }\textbf
  {\bibinfo {volume} {2006}},\ \bibinfo {pages} {P03018} (\bibinfo {year}
  {2006})}\BibitemShut {NoStop}%
\bibitem [{\citenamefont {Berthier}\ and\ \citenamefont
  {Reichman}(2023)}]{berthier2023review}%
  \BibitemOpen
  \bibfield  {author} {\bibinfo {author} {\bibfnamefont {L.}~\bibnamefont
  {Berthier}}\ and\ \bibinfo {author} {\bibfnamefont {D.~R.}\ \bibnamefont
  {Reichman}},\ }\href {\doibase 10.1038/s42254-022-00548-x} {\bibfield
  {journal} {\bibinfo  {journal} {Nature Reviews Physics}\ }\textbf {\bibinfo
  {volume} {5}},\ \bibinfo {pages} {102} (\bibinfo {year} {2023})}\BibitemShut
  {NoStop}%
\bibitem [{\citenamefont {Brooks-Bartlett}\ \emph {et~al.}(2014)\citenamefont
  {Brooks-Bartlett}, \citenamefont {Banks}, \citenamefont {Jaubert},
  \citenamefont {Harman-Clarke},\ and\ \citenamefont
  {Holdsworth}}]{brooks14prx}%
  \BibitemOpen
  \bibfield  {author} {\bibinfo {author} {\bibfnamefont {M.~E.}\ \bibnamefont
  {Brooks-Bartlett}}, \bibinfo {author} {\bibfnamefont {S.~T.}\ \bibnamefont
  {Banks}}, \bibinfo {author} {\bibfnamefont {L.~D.~C.}\ \bibnamefont
  {Jaubert}}, \bibinfo {author} {\bibfnamefont {A.}~\bibnamefont
  {Harman-Clarke}}, \ and\ \bibinfo {author} {\bibfnamefont {P.~C.~W.}\
  \bibnamefont {Holdsworth}},\ }\href {\doibase 10.1103/PhysRevX.4.011007}
  {\bibfield  {journal} {\bibinfo  {journal} {Phys. Rev. X}\ }\textbf {\bibinfo
  {volume} {4}},\ \bibinfo {pages} {011007} (\bibinfo {year}
  {2014})}\BibitemShut {NoStop}%
\bibitem [{\citenamefont {Niggemann}\ \emph {et~al.}(2023)\citenamefont
  {Niggemann}, \citenamefont {Iqbal},\ and\ \citenamefont
  {Reuther}}]{NiggemannPRL2023}%
  \BibitemOpen
  \bibfield  {author} {\bibinfo {author} {\bibfnamefont {N.}~\bibnamefont
  {Niggemann}}, \bibinfo {author} {\bibfnamefont {Y.}~\bibnamefont {Iqbal}}, \
  and\ \bibinfo {author} {\bibfnamefont {J.}~\bibnamefont {Reuther}},\ }\href
  {\doibase 10.1103/PhysRevLett.130.196601} {\bibfield  {journal} {\bibinfo
  {journal} {Phys. Rev. Lett.}\ }\textbf {\bibinfo {volume} {130}},\ \bibinfo
  {pages} {196601} (\bibinfo {year} {2023})}\BibitemShut {NoStop}%
\bibitem [{\citenamefont {Yan}\ \emph {et~al.}(2023{\natexlab{a}})\citenamefont
  {Yan}, \citenamefont {Benton}, \citenamefont {Moessner},\ and\ \citenamefont
  {Nevidomskyy}}]{yan_classification_short}%
  \BibitemOpen
  \bibfield  {author} {\bibinfo {author} {\bibfnamefont {H.}~\bibnamefont
  {Yan}}, \bibinfo {author} {\bibfnamefont {O.}~\bibnamefont {Benton}},
  \bibinfo {author} {\bibfnamefont {R.}~\bibnamefont {Moessner}}, \ and\
  \bibinfo {author} {\bibfnamefont {A.~H.}\ \bibnamefont {Nevidomskyy}},\
  }\href@noop {} {\enquote {\bibinfo {title} {Classification of classical spin
  liquids: Typology and resulting landscape},}\ } (\bibinfo {year}
  {2023}{\natexlab{a}}),\ \Eprint {http://arxiv.org/abs/arXiv:2305.00155}
  {arXiv:2305.00155} \BibitemShut {NoStop}%
\bibitem [{\citenamefont {Yan}\ \emph {et~al.}(2023{\natexlab{b}})\citenamefont
  {Yan}, \citenamefont {Benton}, \citenamefont {Nevidomskyy},\ and\
  \citenamefont {Moessner}}]{yan_classification_long}%
  \BibitemOpen
  \bibfield  {author} {\bibinfo {author} {\bibfnamefont {H.}~\bibnamefont
  {Yan}}, \bibinfo {author} {\bibfnamefont {O.}~\bibnamefont {Benton}},
  \bibinfo {author} {\bibfnamefont {A.~H.}\ \bibnamefont {Nevidomskyy}}, \ and\
  \bibinfo {author} {\bibfnamefont {R.}~\bibnamefont {Moessner}},\ }\href@noop
  {} {\enquote {\bibinfo {title} {Classification of classical spin liquids:
  Detailed formalism and suite of examples},}\ } (\bibinfo {year}
  {2023}{\natexlab{b}}),\ \Eprint {http://arxiv.org/abs/arXiv:2305.19189}
  {arXiv:2305.19189} \BibitemShut {NoStop}%
\end{thebibliography}%



\end{document}

\title{Supplementary Material to\\``Ising Fracton Spin Liquid on the Honeycomb Lattice''}
\author{Benedikt Placke}\email{placke@pks.mpg.de}
\author{Owen Benton}\email{benton@pks.mpg.de}
\author{Roderich Moessner}\email{moessner@pks.mpg.de}
\affiliation{Max Planck Institute for the Physics of Complex Systems, Noethnitzer Str. 38, 01187 Dresden, Germany}

\date{\today}

\begin{abstract}
In this supplementary material, we provide additional details of the cluster Monte-Carlo algorithm used in the main text and show some additional results on the correlations.
\end{abstract} 

\maketitle
\tableofcontents

\section{Detailed Description of the Cluster Algorithm\label{sec:algo}}

\begin{figure}
	\centering\includegraphics{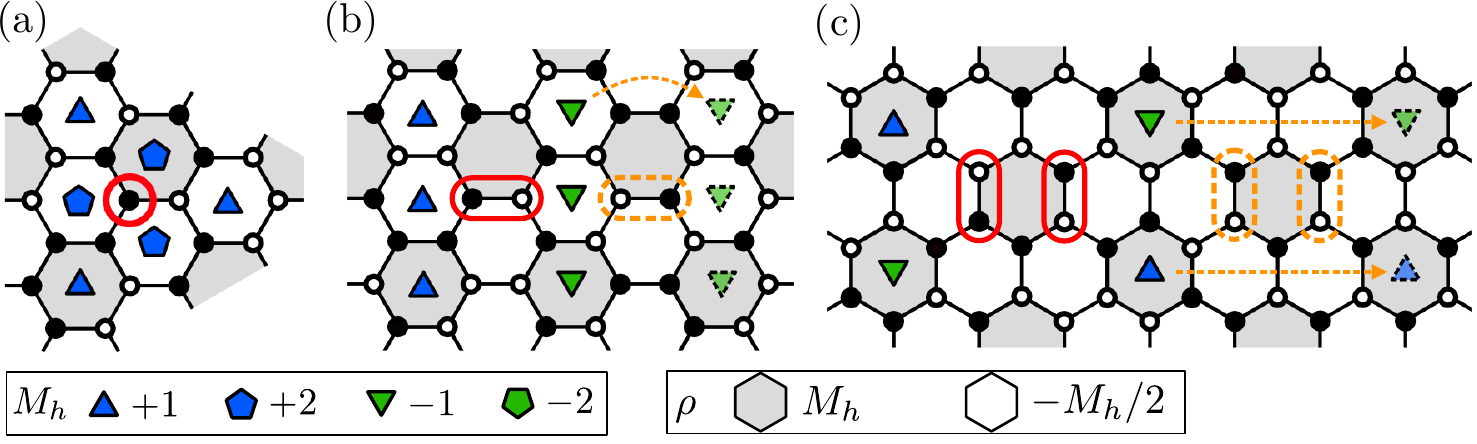}
	\caption{Excitations of the honeycomb lattice model. (a) A single spin flip  with respect to the ground state (indicated by a red circle) creates six violations of the constraint, three single ($M_h=\pm1$) and three double ($M_h=\pm2$) excitatons. (b) Flipping an antiferromagnetic bond creates six single excitations. Flipping also the next bond along the bond direction (encircled in dashed, orange) moves three of these excitations as indicated by the dashed orange arrows. (c) flipping two bond across a hexagon creates four single excitations (fractons), which is the lowest-energy excitation possible. Flipping also the next-next hexagon, now perpendicular to the bond direction, moves two excitations as indicated again in orange.}
	\label{fig:moves}
\end{figure}

As explained in the main text, fractons have restricted mobility. They can only move in pairs, and a fixed pair can only move in one fixed direction. Furthermore, moving a pair of fractons that is a distance $d$ apart also necessitates flipping $\order{d}$ spins and hence it constitutes a nonlocal update of the spin configuration. In the following, we present a cluster Monte-Carlo update rule fulfilling detailed balance which is able to move general pairs of fractons by one step in their direction of mobility.

In particular, we will introduce two kind of cluster moves, which we call the \emph{parallel} and \emph{perpendicular line update}. These are directly inspired by the creation of fractons by flipping membranes of spins with respect to a ground state as shown in \autoref{fig:moves} (b) and (c), respectively. These updates are performed as follows:
\begin{enumerate}
	\item Chose a random bond $b_0=\expval{ij}$. Initialize the cluster as $\mathcal C_0 = \{b_0\}$.
	\item \label{enum:flip} Attempt to flip the cluster $\mathcal C$ with metropolis probability $p=\exp(-\beta \Delta E)$, where $\Delta E$ is the energy difference when flipping all spins in $\mathcal C$. If updated is accepted, go to \ref{enum:accept}.
	\item \label{enum:cluster} If update was rejected add another bond $b'$ to the cluster:
	\begin{enumerate}
		\item If performing the parallel line update, this bond will be the next bond in the lattice reached by going in parallel to the last bond added to the cluster, see \autoref{fig:moves} (b). 
		\item If performing a perpendicular update, this bond will be either the next bond reached when going perpendicular to the cluster if the number of bonds in $C$ is odd, or the next-nearest bond in the same direction, when the number of bonds in $C$ is even, see \autoref{fig:moves} (c).
	\end{enumerate}
	Then go back to \ref{enum:flip}.
	\item \label{enum:accept} Accept the full cluster update with acceptance probability 
	\begin{equation}
		p_{\rm acc} = \prod_{n=1}^{M-1} \frac{1-\min[1, e^{-\beta(\Delta E_{M-n} - \Delta E_M)}]}{1-e^{-\beta \Delta E_n)}}
		\label{eq:fractons:p-accept}
	\end{equation}
	where $\Delta E_n$ is the energy difference in the n-th iteration of the loop above and $M$ is the number of total iterations, that is the size of the cluster $\mathcal C$.
\end{enumerate}

The above procedure implements exactly the movement of fractons as shown in \autoref{fig:moves}, with the parallel line move corresponding to panel (b) and the perpendicular move corresponding to panel (c). 
The cluster $C$ constructed in step \ref{enum:cluster} can be interpreted as one row of a flipped membrane. At low temperature, such a strip will be flipped if either its endpoints coincide with the position of a configuration of fractons, or if it wraps around the system. The former case will move a pair of defects, while the latter case changes the ground state sector. 
In order to ensure that the construction of the cluster always terminates, we remove a bond from the cluster if it would be added a second time. Then, a cluster that wraps around the system twice is empty and hence $\Delta E = 0$.

The acceptance probability in \autoref{eq:fractons:p-accept} accounts for the probability of choosing a certain cluster and is needed to ensure detailed balance. For a in-depth discussion, see for example the derivation of the Wolff algorithm in Sec. 1.6 of Ref. \onlinecite{krauthMonteCarlo}.

\section{Performance of the Cluster Algorithm}

\begin{figure}
\centering
\includegraphics{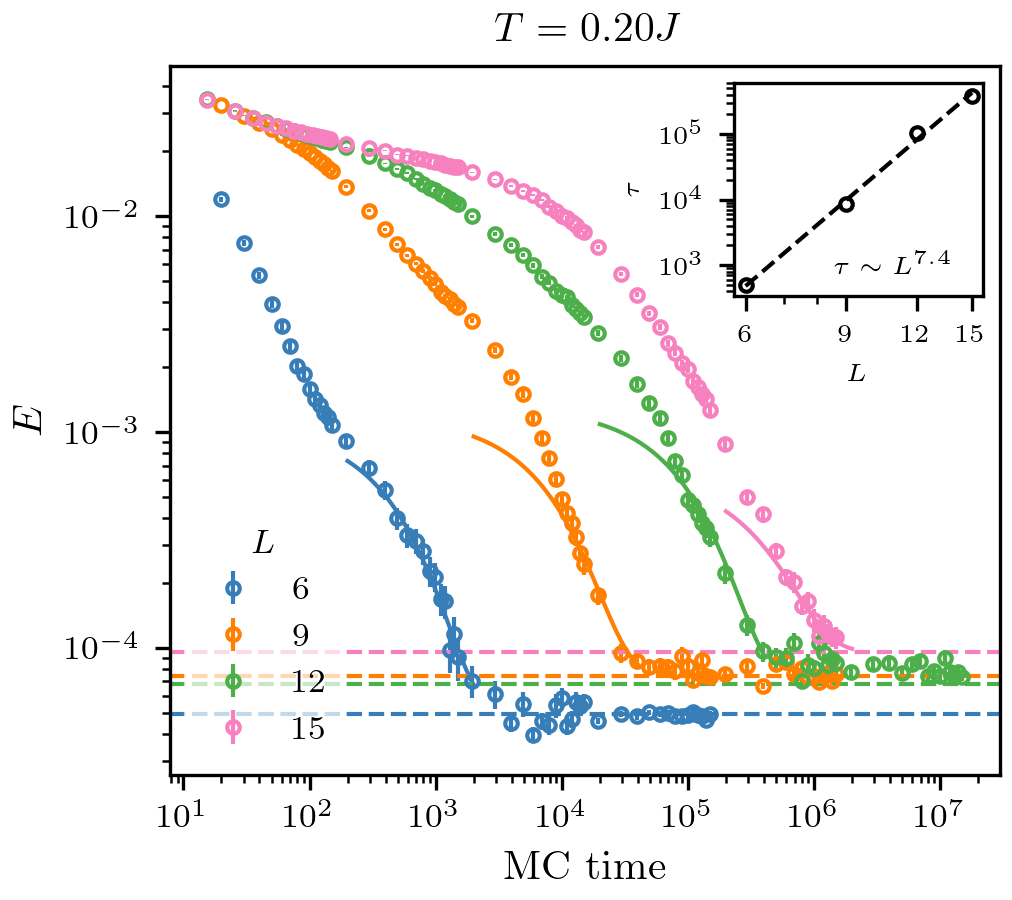}
\caption{Evolution of the internal energy when starting from a random initial state in the ground state regime $T=J/5$. Dashed lines indicate the results from parallel tempering simulations (see main text). Fitting an exponential tail $E-E_{\rm eq} = \exp(-t_{\rm MC}/\tau)$ to the internal energy (solid lines) yields a system size dependence $\tau\sim L^{7.4}$ as shown in the inset.}
\label{fig:quench}
\end{figure}

\begin{figure}
\centering
\includegraphics{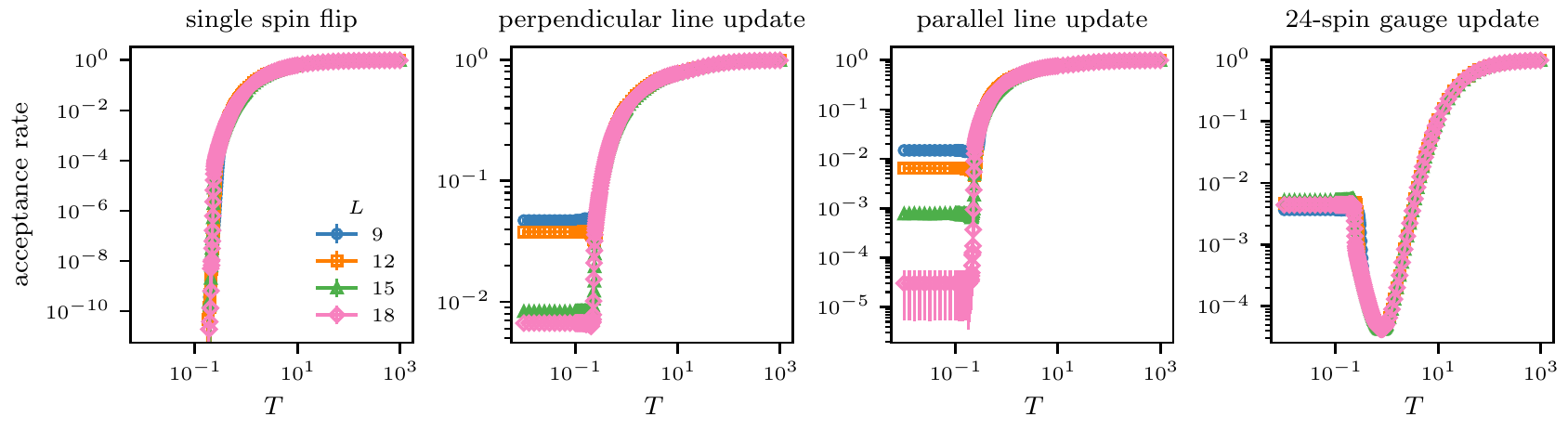}
\caption{Acceptance rates of the difference mocves during simulated annealing as a function of temperature. Here ``vertex move'' denotes single-spin flips and ``gauge move'' denotes the 24-spin flip shown in Fig. 2 (b,c) in the main text, both accepted with metropolis probability. The ``line moves'' are the two types of cluster moves as described in \autoref{sec:algo}. 
All results are averaged over 500 annealing runs.}
\label{fig:acceptance}
\end{figure}

\subsection{Relaxation dynamics after a quench}

To benchmark our algorithm and as a consistency check of ergodicity, we show in \autoref{fig:quench}, the internal energy of our model as a function of Monte-Carlo time, when starting from a random initial state at a temperature of $T=J/5$. This temperature is well in the ground state regime, with much less than one excitation present on average. The data between times $T_{\rm MC}=10^m$ and $T_{\rm MC}=1.5 \times 10^{m+1}$ is obtained by alternating $9 \times 10^{m}$ Monte-Carlo sweeps with $\times 10^{m}$ measurements and repeating this $15$ times. The data is averaged over 5000 runs for $L=6, 9$ and over 2000 runs for $L=12, 15$.
We compare the long time limit of this quench dynamics with the results from parallel tempering simulations (see main text). While the long term limits show reasonable agreement with the simulated annealing data within estimated errors, the equilibration time seems to diverge strongly with system size. 
Still, the cluster algorithm poses a major improvement over any local version, which fails to reach the equilibrium value starting from a random initial state within $10^7$ sweeps for all system sizes shown.

\subsection{Acceptance rates as a function of temperature}
The superiority of the cluster algorithm is also well illustrated by inspecting the acceptance probabilities of the difference types of moves as a function of temperatures as shown in \autoref{fig:acceptance}. In the low temperature regime, single spin flips almost never get accepted (virtually never during our simulation), while the cluster moves as well as the ``gauge move'' (meaning the 24-site minimal local move shown in Fig 2 (b,c) in the main text) have low, but nonzero acceptance probabilities even in the low-temperature ground state regime.

\section{Additional data from Monte-Carlo simulations}

\subsection{Structure factor as a function of temperature}

\begin{figure}
    \centering{}
    \includegraphics{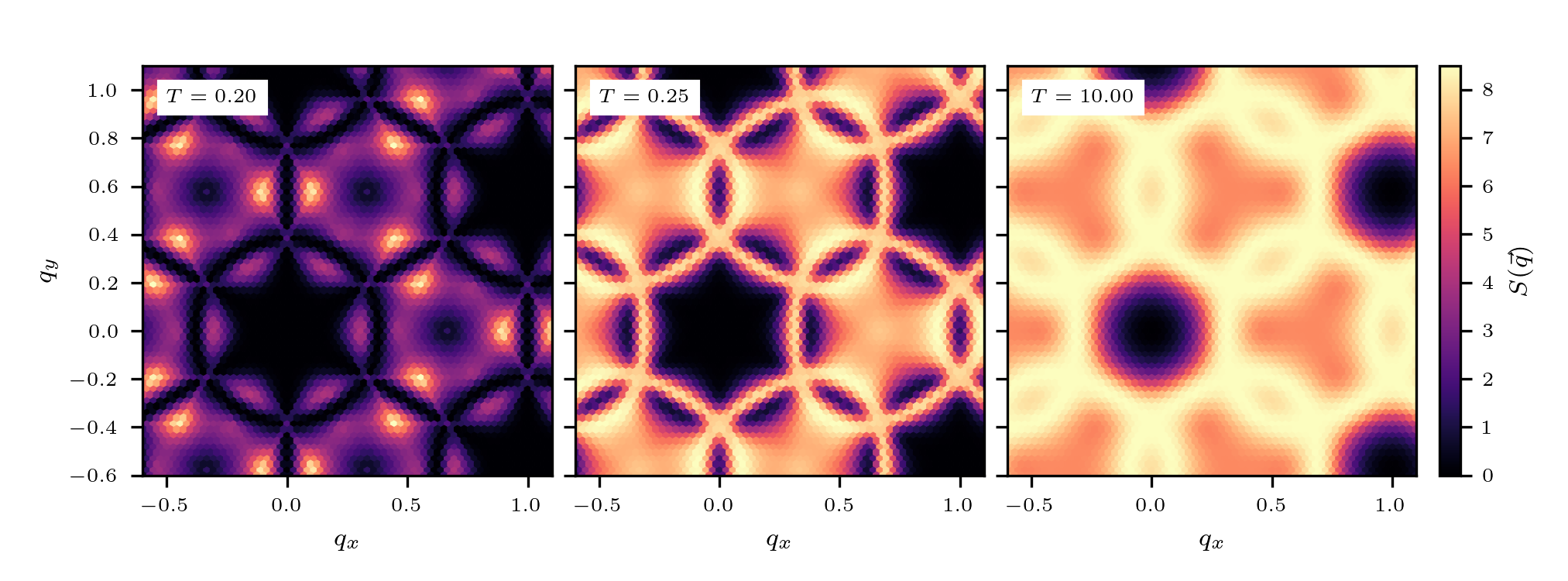}
    \caption{Structure factor as a function of temperature for system size $L=24$. While the onset of the four-fold structure at the Brillouin-zone corners is already visible above the transition ($T=J/4$), the pinch points only sharpen below the transition ($T=J/5$)}
    \label{fig:sq}
\end{figure}

In \autoref{fig:sq}, we show the structure factor as a function of temperature for a system with $L=24$. Above the ground state regime, at $T=J/4$, the four fold pinch points are already beginning to form, but only become sharp after the first-order transition.

\subsection{Higher-order correlations}

\begin{figure}
    \centering{}
    \includegraphics{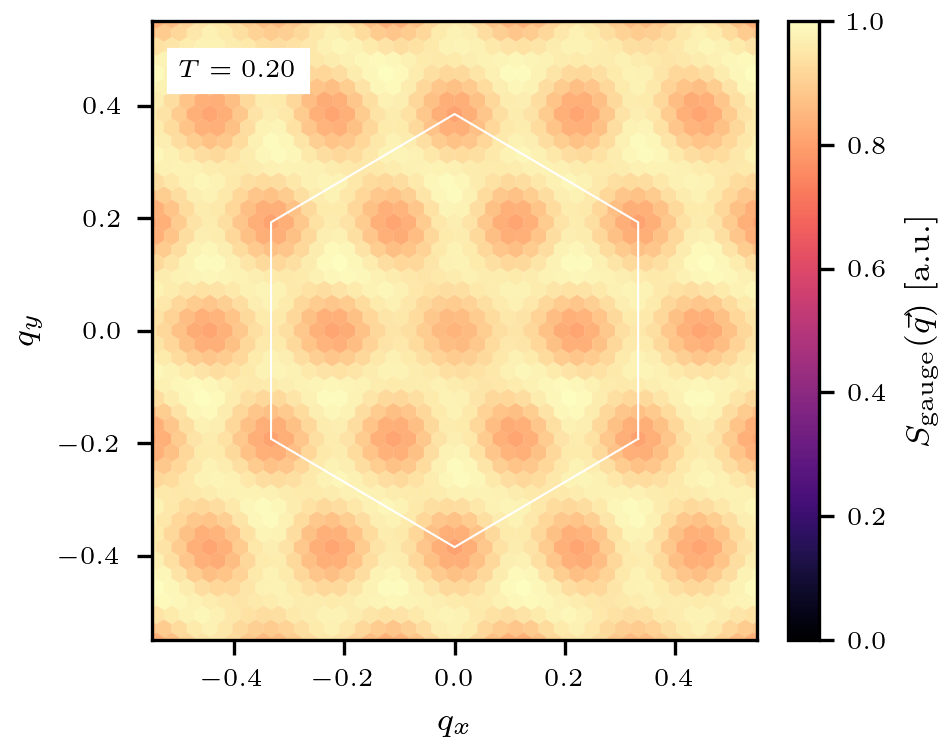}
    \caption{Gauge structure factor at low temperature, corresponding to the Fourier transform of the correlation function of the 24-site local motif shown in Fig 2 (b,c) in the main text. The plot is the result of a parallel tempering simulation of a $L=24$ system.}
    \label{fig:gauge-sq}
\end{figure}

As discussed in the main text, an open question in the absence of an exact solution is the degree to which the gorund state manifold breaks symmetry. There are no strong indicators for symmetry breaking in the two-point correlation function (see structure factor shown in Fig. 4 of the main text). However, in a system with such unusual constraints one can imagine also order showing up only --- or at least more prominently ---  in higher-order correlation functions. While an exhaustive study of those is obviously not possible, a good intuition for order through entropy is often obtained from correlations corresponding to the local moves connecting different ground states. To this end, we show in \autoref{fig:gauge-sq} what we call the ``gauge structure factor'' in the low temperature regime for a system size of $L=24$. This quantity if defined as the Fourier transform of the correlation function between flippable gauge motifs. That is for each hexagon, we define a ``gauge flippability'' as
\begin{equation}
    g(h) = \begin{cases}
        0 & \text{gauge flip centered at $h$ costs nonzero energy} \\
        \pm 1 & \text{else, sign depending on the current configuration}
    \end{cases}
\end{equation}
and accordingly
\begin{equation}
    S_{\rm gauge} (\vec q) \sim \sum_{h, h'} \exp(-i \vec q \cdot (\vec r_{h} - \vec r_{h'})) \expval{g(h)g(h')}.
\end{equation}

While $g(h)$ quantity shows short range correlations as evidenced by the structure factor in \autoref{fig:gauge-sq}, we find no evidence for long-range correlations at this system size.

\bibliography{fractons}